\documentstyle[prd,aps]{revtex}
\begin{document}
\draft
\input epsf
\twocolumn[\hsize\textwidth\columnwidth\hsize\csname
@twocolumnfalse\endcsname

\title{Hybrid Inflation in Supergravity}
\author{Andrei Linde$^{(1)}$ and Antonio
Riotto$^{(2)}$}
 \address{$^{(1)}${\it Department of Physics, Stanford University,
Stanford, California~~94305-4060}}
\address{$^{(2)}${\it NASA/Fermilab Astrophysics Center, \\ Fermilab
National Accelerator Laboratory, Batavia, Illinois~~60510-0500}}
\date{March 2, 1997}
\maketitle
\begin{abstract}
We propose a simple realization of  hybrid inflation  in supergravity, which does not require any unnaturally small parameters. Inflation in this scenario is relatively short, and inflationary  density perturbations   fall down on   large scales, which corresponds to a blue spectrum with  $n > 1$.  Superheavy strings which are formed after inflation in this scenario produce density perturbations of comparable magnitude. In this mixed perturbation (MP) model one may have microwave anisotropy produced by strings and galaxy formation due to inflationary perturbations.
\end{abstract}
\pacs{PACS: 98.80.Cq    \hskip 1.2 cm FERMILAB--Pub--97/051-A~~~~~~
SU-ITP-97-07 \hskip 1.4 cm
 hep-ph/9703209}
\vskip2pc]

For the last 15 years cosmologists   believe  that something like
inflation is
necessary in order to construct an internally consistent cosmological theory.
For more than 20 years the best hopes for the theory of all fundamental
interactions are related to supersymmetry. Unfortunately it is not that simple
 to
combine these two ideas.

The main problem is that the effective potential for the inflaton field
$\sigma$ in supergravity typically is too curved, growing as
$\exp{C\sigma^2\over M^2}$ at large values of the inflaton field $\sigma$. Here $M   = 2.4\times 10^{18}$
GeV  is the stringy   Planck mass. The typical value of the parameter $C$ is
$O(1)$, which makes inflation impossible because the inflaton
mass  in this theory becomes  of the order of the Hubble constant. Typical
shape of the scalar field potential in superstring theory is $\exp{C\sigma\over
M}$, which prevents inflation for $\sigma > M$.

There are some cases where successful inflationary
scenario can be developed despite these problems.
Perhaps the easiest way to find a consistent inflationary
model in
supersymmetric theories is based on the hybrid inflation scenario
\cite{Hybrid}.  In particular, recently a very interesting class of hybrid
inflation models have
been proposed based on the use of the D-term in the effective
potential, which does not contain the exponentially growing factors \cite{BDH}.
However, in this paper we will concentrate on the   model with the   simplest
choice of the superpotential which leads to hybrid
inflation   \cite{Cop}
 \begin{equation}\label{1}
W=  S (\kappa\bar\phi\phi - \mu^2),
\end{equation}
 with $\kappa \ll 1$. Here $\phi$ and $\bar{\phi}$ denote conjugate pairs of
superfields transforming as non-trivial representations of some gauge group $G$
under which the superfield $S$ is neutrally charged. As noted in \cite{dvali},
the superpotential (\ref{1}) is of the most general form consistent with an
$R$-symmetry under which $S\rightarrow {\rm e}^{i\alpha}S$, $W\rightarrow {\rm
e}^{i\alpha}W$ and $\bar\phi\phi$ invariant. Hybrid   inflation scenario with
this superpotential in a context of a globally supersymmetric theory was
studied by many
authors, see e.g.  \cite{Cop,dvali,jeannerot}. However,   the possibility to
have inflation in this model with an account taken of supergravity corrections   was not fully
investigated. This will be the purpose of the present paper.

The effective potential in a globally supersymmetric theory with the
superpotential (\ref{1}) is given by
\begin{equation}\label{2}
V =  {\kappa^2|\sigma|^2\over 2}\bigl(|\phi|^2 +|\bar\phi|^2\bigr)
+|\kappa\bar\phi\phi -\mu^2|^2 +   D-{\rm term  }.
\end{equation}
Here    $\sigma = \sqrt 2 S$   is   a canonically
normalized scalar field.
The absolute minimum appears at  $\sigma = 0$, $\phi = \bar\phi = {\mu\over
\sqrt
\kappa}$. However, for $\sigma>\sigma_c  = {\sqrt 2\mu\over \sqrt \kappa}$, the
fields $\phi$ and $\bar\phi$ possess a positive mass squared and stay at
the origin.
This potential for $\phi = \bar\phi = 0$ is exactly flat in the
$\sigma$-direction. If one simply adds a mass term $ {m}^2\sigma^2/2$ which
softly breaks supersymmetry, one obtains a simple realization of the hybrid
inflation scenario \cite{Cop}: Initially the scalar field $\sigma$ is large. It
slowly rolls down to $\sigma = \sigma_c$. Then the curvature of the effective
potential in the $\phi$-direction becomes negative, the fields rapidly roll to
the absolute minimum of the effective potential leading to the symmetry
breaking of the group $G$, and inflation ends, as in the
original version of the hybrid inflation scenario \cite{Hybrid}.

However,   if one takes into account radiative corrections and supergravity
effects, the behavior of the fields becomes somewhat different. The one-loop
effective potential in this model is easily calculated  from the spectrum of
the theory composed by  two pairs of real and pseudoscalar fields with mass
squared $\kappa^2
 \sigma^2/2 \pm \kappa\mu^2$ and a Dirac fermion with mass $ \kappa
\sigma/\sqrt 2$. The one-loop effective potential is given by
\cite{dvali}
\begin{eqnarray}\label{3}
V_{1}&=& \frac{\kappa^2}{128 \pi^2} \Biggl[(\kappa \sigma^2 - 2\mu^2)^2 \ln{
\kappa \sigma^2 - 2\mu^2\over  \Lambda^2}\nonumber\\& +&  (\kappa \sigma^2 +
2\mu^2)^2 \ln{\kappa \sigma^2 + 2\mu^2\over  \Lambda^2} -  2   \sigma^4 \ln
{\kappa \sigma^2\over  \Lambda^2} \Biggr],
\end{eqnarray}
where $\Lambda$ indicates the renormalization scale.

At the stage of  inflation, when    $\sigma\gg \sigma_c$,  the total effective
potential is
\begin{equation}\label{XX}
V = \mu^4\Bigl[1 + {\kappa^2\over 8\pi^2} \ln { \sigma\over \sigma_c} +...
\Bigr] \ .
\end{equation}
The Hubble constant practically does not change during inflation. In units $M = 1$, which we will use throughout the paper, one has
$H =  \mu^2/\sqrt{  3} .$ 

It is convenient to use time $t$ measured in units of $H^{-1}$. In these units time $t$ is directly related  to the number of e-folds $N$. The usual equation $3H\dot\sigma = -V'$ for the field $\sigma$ in this system of units looks particularly simple:
\begin{equation}\label{4}
 \dot \sigma = -{V'\over V}  = -  {\kappa^2  \over 8 \pi^2\sigma}.
\end{equation}
This gives
 \begin{equation}\label{6}
\sigma_0^2 - \sigma^2(t) = {\kappa^2   \over 4\pi^2 }  t.
\end{equation}
This means that the universe expands during the inflationary stage as follows:
\begin{equation}\label{7}
a(t) = a(0)e^{t}= a(0) \exp\Bigl( {4\pi^2\over
\kappa^2 } (\sigma_0^2 - \sigma^2(t))\Bigr).
\end{equation}
The total number of e-folds of  inflation is obtained by taking $\sigma(t) = 0$:
\begin{equation}\label{7a}
N = \ln {a(t) \over  a(0)} \approx  {4\pi^2  \sigma_0^2  \over \kappa^2   }.
\end{equation}
The universe expands $e^N$ times when the field $\sigma$ rolls down
from 
\begin{equation}\label{SIGMAN}
\sigma_N = {\kappa \sqrt N\over 2\pi}.
\end{equation}
In particular, the structure of the observable part of the universe,
corresponding to $N \sim 60$,  is formed   at  $\sigma_{60} \sim 1.2 \kappa    \ll
1$.

Now let us calculate density perturbations \cite{book}:
\begin{equation}\label{8}
{\delta\rho\over\rho} \sim {\sqrt 3\over 5\pi}   {V^{3/2}\over V'  } =
{8\pi\sqrt 3\over 5\kappa^2} {\mu^2 \sigma }.
\end{equation}
Using (\ref{7a}) one can express this result   in terms of the ratio  of the
wavelength  $l$ of a perturbation to the   wavelength $l_0$ of a perturbation
which had the wavelength $H^{-1}$ at the end of inflation:
\begin{equation}\label{8a}
{\delta\rho\over\rho} \sim
{4\sqrt 3 {\mu^2 }\over 5\kappa }  \ln ^{1/2}{ l\over l_0 }.
\end{equation}
The COBE normalization is \cite{Lyth}
\begin{equation}\label{COBE}
 {V^{3/2}\over V'  }= {8\pi^2\over  \kappa^2} {\mu^2 \sigma } \sim
5.3 \times 10^{-4}.
\end{equation}
For $\sigma \sim 1.2 \kappa $ it gives $ {\mu   }\sim 2.5 \times 10^{-3}\, \sqrt\kappa$\,  (in units $M = 1)$. This can be achieved {\it e.g}. for  $\kappa \sim  0.1$, and $\mu \sim 2\times
10^{15}$ GeV; we will make a more accurate estimate shortly.  Note that after the symmetry breaking in this scenario one may
encounter cosmic strings production on a very interesting mass scale $\phi \sim
{\mu\over\sqrt \kappa} \sim 10^{16}$ GeV. This may happen, for
instance,  during the second stage of the symmetry breaking   of the
supersymmetric $SO(10)$ GUT group, $SO(10)\rightarrow SU(3)_c\otimes
SU(2)_L\otimes U(1)_R\otimes U(1)_{B-L}\rightarrow SU(3)_c\otimes
SU(2)_L\otimes U(1)_Y\otimes Z_2$. Even though during the first stage of this
symmetry breaking topologically stable monopoles can be formed, they are
subsequently diluted at the inflationary stage preceding the phase transition
when $\phi$ and $\bar\phi$ acquire a vacuum expectation value and
$SU(3)_c\otimes SU(2)_L\otimes U(1)_R\otimes U(1)_{B-L}$ breaks down to
$SU(2)_L\otimes U(1)_Y\otimes Z_2$. In such a case
the $\phi+\bar\phi$ have to be identified with a $16+\overline{16}$ or a
$126+\overline{126}$ pair of Higgs fields and $S$ is an $SO(10)$ singlet \cite{jeannerot}.  As we will see, inflation ends well before this second phase transition takes
place and therefore the cosmic strings  formed when   $\phi$ and $\bar\phi$
roll down to their true values are not diluted.

It is important to understand that radiative corrections change the way
inflation ends in the hybrid inflation model. In the original version of this
  scenario the structure of the universe appears due to the perturbations generated at  at $\sigma \sim \sigma_c = {\sqrt 2
\mu\over \sqrt\kappa}$  \cite{Hybrid}. For  $\kappa \sim  0.1$, and $\mu \sim
2\times 10^{15}$ GeV one has $\sigma_c \sim 10^{16}$ GeV. Meanwhile in our
scenario this happens for $\sigma_{60} \sim 1.2 \kappa    \ll
1$.
For $\kappa \sim  0.1$ one has $\sigma_{60} \sim 3\times 10^{17}$ GeV$\sim 30
\sigma_c$. Therefore when one calculates density perturbations and evaluates a possible significance of SUGRA
corrections to the effective potential in our scenario, one should do it not
near $\sigma_c$, but at $\sigma \sim k M \gg \sigma_c$.  

The  supergravity potential for $\phi = \bar\phi = 0$ (ignoring the one-loop
corrections calculated above) is given by \cite{Cop}
\begin{eqnarray}\label{SUGRAPOT}
V_{\rm SUGRA} &=&  \mu^4 \exp \Bigl({\sigma^2\over 2 }\Bigr) \Bigl[1-
{\sigma^2\over 2 } + {\sigma^4\over 4 }  \Bigr]\nonumber\\
&= & \mu^4   \Bigl[1 + { \sigma^4\over 8 }+ ...  \Bigr].
\end{eqnarray}
Inflation with  this potential is possible because of the cancellation of the
quadratic term   $\sim {\mu^4 }\sigma^2=3H^2\sigma^2$, but still it may
occur only at $\sigma
\lesssim 1$. Notice that the cancellation of this quadratic term derives from
the general  form of the typical superpotential during hybrid inflation, $W\sim
\mu^2 S$. However, this cancellation is only operative with the choice of
minimal K\"{a}hler potential $K=S^{\dagger}S$.

Since the K\"{a}hler potential is not protected by any nonrenormalization
theorem, it is reasonable to expect the presence of terms like \begin{equation}
 \delta{\cal L}=\lambda \int d^4\theta{(S^{\dagger}S)^2}
=  2   \lambda \mu^4 \sigma^2
\end{equation}
in the effective Lagrangian. This mass term, however, is not important for our
consideration if
$\lambda {\mu^4} \sigma< {\rm Max}\{V'_1,V'_{\rm SUGRA}\}$ in the range
 $\sigma_0 <\sigma< \sigma_N$. This translates into the bound
bound $\lambda \lesssim 0.03  {\kappa} $, which is   restrictive but   not   dangerous  if
$\delta{\cal L}$ is generated at the one loop level.

 Also, as we will see now, at small $\sigma$ the one-loop SUSY
potential is more important, so that at the last stages of inflation the
effective potential is always dominated by the one-loop effects.

Indeed, for $\sigma < 1$ the effective potential is approximately given by
$\mu^4$, so it remains to compare its derivative due to one-loop and SUGRA
terms. The condition $V'_1 \gtrsim V'_{\rm SUGRA}$ reads
$\sigma  \lesssim    {\sqrt \kappa \over 3 }.$
Comparing this expression with (\ref{SIGMAN}) one concludes, that the last $N_{\rm SUSY} \sim  {4 \kappa^{-1}}$~ e-folds of inflation are controlled by the loop effects. In
particular, in order to neglect SUGRA effects during the last 60 e-folds of
inflation one would need to have $\kappa \lesssim 6\times 10^{-2}$. This condition
is reasonable, but it is not automatically guaranteed, as   expected in
\cite{dvali}.

Now let us study inflation at the stage when SUGRA effects are dominant. In
this case equation of motion is particularly simple,  ${\dot \sigma } = - \sigma^3/2.$ 
It has a solution     $\sigma^{-2}(t) -  \sigma^{-2}_0 =      t  .$
Suppose that inflation begins at $\sigma_0 = 1$, $t  = 0$. Then  
$ \sigma^{-2}(t)   =  1 +   t$. Expressing everything in terms of the number of e-folds from the beginning of
inflation $N_{\rm SUGRA} =  t$, one has
\begin{equation}\label{SIGSUGRA2}
   \sigma (t)   =   {1\over \sqrt{1  +    N_{\rm SUGRA}}} \approx  {1\over
\sqrt{ N_{\rm SUGRA}}}.
\end{equation}
As we already noted, the SUGRA term dominate only at $\sigma > {\sqrt \kappa \over 3 }$, which gives the total number of e-folds of expansion at the
SUGRA stage: $ N_{\rm SUGRA}  \sim  {9  \kappa^{-1}}$. Thus   inflation consists of two  long stages, one of which is
determined by the one-loop effects, another, which is about two times longer, is determined by SUGRA corrections. In fact, the total duration of each of these stages is slightly shorter than what is indicated by our estimates because at an intermediate stage of inflation the contributions of radiative corrections and SUGRA terms are comparable. This increases the speed of rolling of the field $\sigma$ and makes inflation slightly shorter. The total
duration of inflation can be estimated  by
\begin{equation}\label{SIGSUGRA3}
 N_{\rm SUGRA}+ N_{\rm SUSY}  \sim  {10  \kappa^{-1}}.
\end{equation}
Therefore we need $\kappa \lesssim 0.15$ in order to obtain 60 e-folds of
inflation in our scenario.

During the SUGRA stage the density perturbations have   spectrum
\begin{equation}\label{88}
{\delta\rho\over\rho} \sim {\sqrt 3\over 5\pi}   {V^{3/2}\over V' } = {2
\mu^2  \over 5\pi\sqrt{3} \sigma^3 } \ .
\end{equation}

Using (\ref{SIGSUGRA2}) one can express it in terms of the ratio of the
wavelength $l$ to the wavelength $l_{\rm max}$ corresponding to the beginning
of the first stage of inflation:
\begin{equation}\label{8aa}
{\delta\rho\over\rho} \sim   {6
\mu^2  \over 5\pi   } \,   \ln ^{-3/2}{ {l_{\rm max}\over l} } \ .
\end{equation}

Eqs. (\ref{8a}) and (\ref{8aa}) match each other at the
scale $l^*$ at which SUGRA terms become smaller than the radiative corrections.
This scale therefore should correspond to a maximum of the spectrum
${\delta\rho\over \rho}(l)$.

At $l > l^*$ the spectrum decreases towards large scales, which corresponds to
a blue spectrum. In the beginning, when the logarithm is rather large, the
damping of the amplitude of density fluctuations is relatively insignificant.
However, when $l$ approaches $l_{\rm max}$ the spectrum falls down very
sharply.

To evaluate the change of ${\delta\rho\over\rho}$ during the   SUGRA inflation
(for $\kappa = 0.1$), one should note that in the beginning one has $\sigma
\sim 1$, and in the end of SUGRA inflation one has $\sigma \sim  {\sqrt \kappa
  \over 3 }$. This implies that at the interval from $l^*$ to $l_{\rm max}$
density perturbations
fall down by $27 \kappa^{-3/2} \sim 10^3$. This means that in order to make
this model consistent with COBE data one should keep $l_{\rm max}$ somewhat greater
than the size of the observable part of the universe.

To study this question in a more detailed way we solved equations for the field $\sigma$ numerically, taking into account simultaneously the contribution of SUGRA terms and radiative corrections, assuming in the first approximations that these two types of terms are additive. The results of our investigation are shown in Figs. 1 and 2 for $\kappa = 0.1$.
The total duration of inflation in this case is given by $N \sim 90$.  The spectrum of density perturbation depends on scale $l \sim e^N$. However, near $N \sim 60$ where the COBE normalization should be imposed this dependence is rather mild. It can be expressed in terms of the effective spectral index $n$, which in our case is given by $1+ 2V''$. At $N \sim 60$ one has $n -1 \sim 0.1$, which is a noticeable deviation from the flat Harrison-Zeldovich spectrum. From COBE normalization in that case one finds $\mu \approx 1.2 \times 10^{-3} M \sim 3 \times 10^{15}$ GeV, and the symmetry breaking scale $\phi \sim {\mu\over \sqrt\kappa} \sim    10^{16}$ GeV.

\begin{figure}[t]
\centering
\leavevmode\epsfysize=4cm \epsfbox{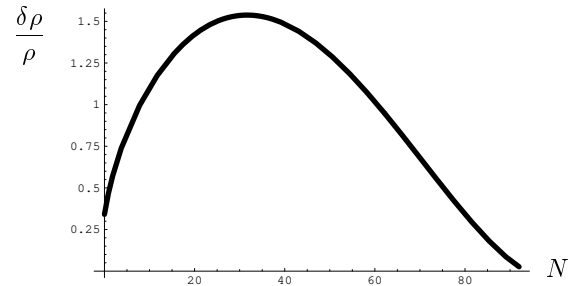}\\
\caption[fig1]{\label{fig1} The shape of  density perturbations as a function of the number of e-folds  for $\kappa = 0.1$.  The horizon scale corresponds to $N \sim 60$. COBE normalization in this figure is shown as ${\delta\rho\over\rho} = 1$. Density perturbations sharply fall down on the scale corresponding to the beginning and the end of inflation.   }
\end{figure}

For $\kappa < 0.05$ the total duration of inflation is $N \sim 200$, and the last 60 e-folds are determined by the one-loop effects. In that case  the effective spectral index  at the horizon scale is very close to 1,   $\mu$ is two times smaller than for  $\kappa < 0.1$, and symmetry breaking scale is $\phi \sim 7\times   10^{15}$ GeV. For $\kappa$ approaching 0.15  the total duration of inflation approaches $N \sim 60$. In this case the spectrum of density perturbation   sharply falls down on the horizon scale, which corresponds to $n \gg 1$, in contradiction with observations. Finally, for $\kappa > 0.15$ inflation is too short to incorporate the observable part of the universe.
\begin{figure}[t]
\centering
\leavevmode\epsfysize=4cm \epsfbox{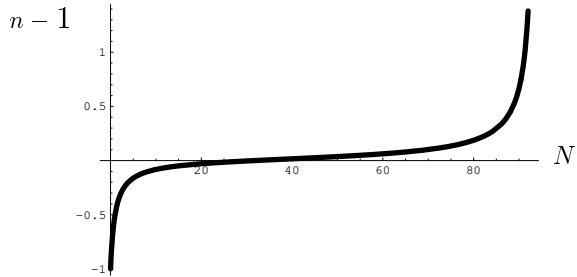}\\
\caption[fig2]{\label{fig2} Deviation of the effective spectral index from $n =1$ at different scales for $\kappa = 0.1$. For    $\kappa > 0.15$ the duration of inflation becomes smaller than $60$ e-folds, and the spectral index rapidly grows near the horizon.}
\end{figure}

This behavior is rather interesting and unusual. Most of the inflationary models   describe
 the universe with a long stage of inflation, and with a spectrum slowly
growing on large scale. Hybrid inflation provided a  natural way  to
obtain a blue spectrum of perturbations, decreasing on large scale. However, in
the simplest versions of this scenario the spectrum was nearly flat. Recently a
new class of models was introduced, which was called tilted hybrid inflation
\cite{BELL}. In such models one may have a rather short stage of inflation and
blue spectrum of perturbations. Both of these features may be useful for
constructing realistic models of open inflationary universe \cite{BELL2}. It
was not quite clear, however, whether the models with such properties could be
constructed in the context of a realistic theory of elementary particles. It is
very interesting (and even somewhat unexpected) that when one makes an attempt
to implement hybrid inflation scenario in the context of a
supersymmetric model with the simplest superpotential (\ref{1}), one obtains
the hybrid
inflation scenario  of this new type.

Until now we ignored density perturbations produced by strings formed after inflation.  These perturbations are proportional to $\phi^2\sim \mu^2/\kappa$. According to \cite{jeannerot},   stringy perturbations in this model  are expected to be of the same order as inflationary ones, though a little smaller. Our results taking into account supergravity corrections confirm this conclusion: Strings which are generated in this model appear as as result of spontaneous symmetry breaking with $\phi \sim 10^{16}$ GeV, which is the right scale for stringy density perturbations. The relative importance of the two types of perturbations depends on the choice of parameters in our model. For example,  by decreasing $\kappa$ from $0.1$ to $0.05$ and decreasing  $\mu$ two times one suppresses stringy perturbations by a factor of 2 without changing the inflationary ones. Perturbations of both types   are proportional to $\mu^2$. Therefore keeping all other parameters fixed, one can easily  adjust the combined amplitude of fluctuations to the COBE normalization without affecting much of our results. Finally, one can modify the theory in such a way as to avoid topological defects altogether, see e.g. \cite{smooth}.

We will call the model where where inflationary and stringy perturbations have comparable magnitude {\it a  mixed perturbation} (MP) model. An interesting possibility  which appears in the MP model  is related to the different scale dependence of these  perturbations. Stringy perturbations are approximately scale-independent. If we consider a scenario with a blue spectrum of inflationary perturbations, one may encounter a situation where the perturbations of metric on the galaxy scale are dominated by   adiabatic inflationary perturbations, whereas the perturbations on the horizon scale, which show up in the large angle anisotropy of the microwave background radiation, are dominated by strings. It seems that the hybrid inflation scenario may indeed live  up to its name!

 A.L. is grateful to J. Garc\'{\i}a--Bellido for important comments, and to M. Dine for suggesting to study models with the superpotential (\ref{1}) back    in 1991. A.R. would like to thank D.H Lyth  and W. Kinney for several discussions.    A.L.\ is supported in part by   NSF grant PHY-9219345.    A.R.\ is supported by the DOE and NASA
under grant NAG5--2788. 

After this work was finished we learned about a related
work by  C. Panagiotakopoulos, hep-ph/9702433. The author also studied SUGRA
effects in the model (\ref{1}), but he did not consider a combined scenario
containing both SUGRA terms in the effective potential and radiative
corrections.  Moreover, he calculated density perturbations near $\sigma_c$ and concluded that one needs to have extremely small values of the parameter $\kappa$. As we emphasized, density perturbations should be evaluated  at $\sigma_{60} \gg \sigma_c$, which allows  to have small density perturbations with a reasonably large $\kappa$.

\vskip 3 cm

\end{document}